**White-Light Colour Photography for Rendering Holo-Images in a Difractive Screen**

J.J Lunazzi
Campinas State University – São Paulo – Brazil

 Abstract

 The capability of color-encoding the continuous sequence of views from a scene was demonstrated previously by the author (1990). In the present work, the scheme for this process is shown where white light from a black and white object is diffracted at a diffraction grating and then photographed on colour film. Two rays of different wavelengths reaching the plane of the color photographic film determine the stereo representation of an object point. Since the wavelength may have any value within the continuous visible spectrum, this case constitutes a new situation
of continuous stereo-photography. A natural process of decoding is represented where a diffusing white light source was added from the side of the developed photographic film. One white-light ray that matches the former position of an incident ray receives the spectral characteristics of the registered point when traversing the photographic slide. It characterizes a situation of light path reversal, and the ideal result corresponds to a projected white-light point being focused at
 the original object position. This situation generates a pseudoscopic image of the object, as seen from a certain distance, whose colour depends on the horizontal position of the observer.

**Introduction**

The capability of colour-encoding the continuous sequence of views from a scene was demonstrated in a previous paper by Lunazzi (1). We show the scheme for this process in figure (1a) where white light from a black and white object is diffracted at a diffraction grating and then photographed on colour film. Two rays of different wavelengths reaching the plane of the colour photographic film determine the stereo representation of an object point.

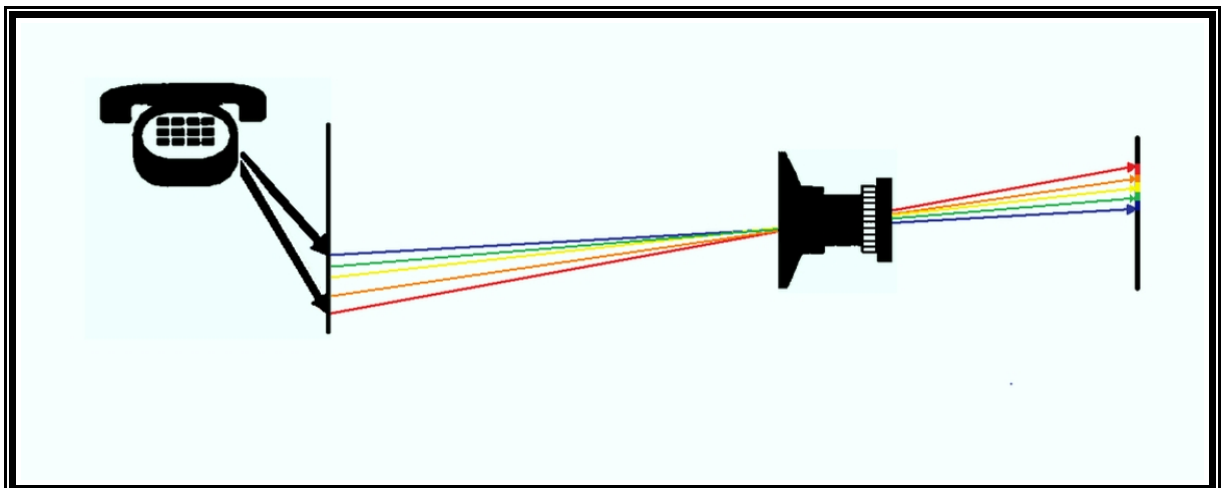

*Figure 1 a : The photographing of a diffracted image.*

White light is represented by wide continuous lines, short wavelength light by thinner continuous lines while a longer wavelength value is represented by dashed lines (in this digital version of the paper colour representation makes the figures more clear). Since the wavelength may have any value within the continuous visible spectrum, this case constitutes a new situation of continuous stereo–photography. A natural process of decoding is represented in figure (1b) where a diffusing white light source was added from the side of developed photographic film. One white light ray that matches the former position of an incident ray receives the spectral characteristics of the registered point when traversing the photographic slide.

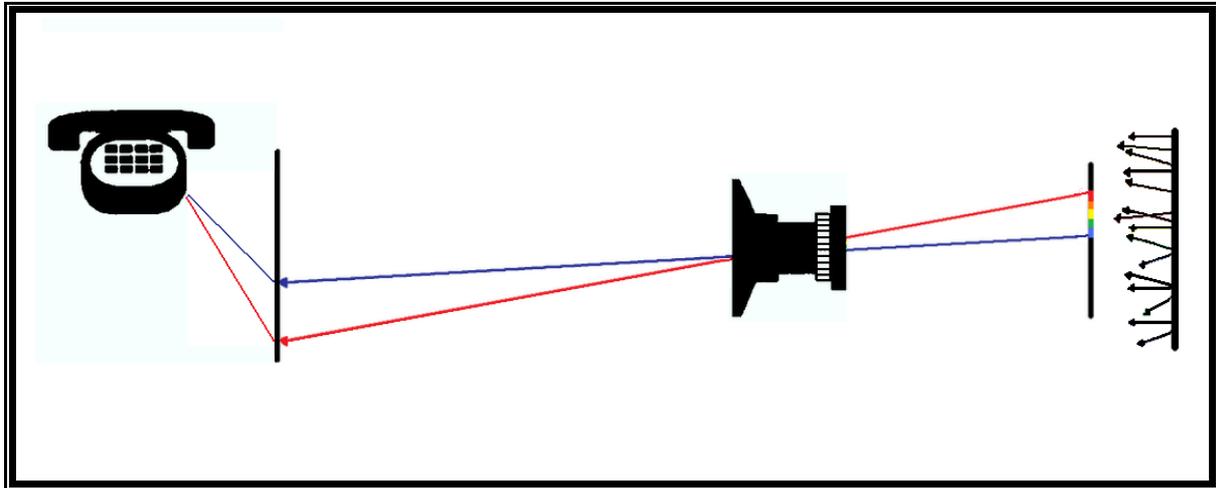
Figure 1b : A diffusing white light source was added from the side of the developed photographic film.

It characterizes a situation of light path reversal, and the ideal result corresponds to a projected white – light point being focused at the original object position. This situation generates a pseudoscopic image of the object, as seen from a certain distance, whose colour depends on the horizontal position of the observer.

**Obtaining an Orthoscopic Image**

The image can be made orthoscopic by two equivalent procedures:

I. Observing the light as diffracted to the symmetrical diffraction order of the grating, as reported by Lunazzi (2).

II. Reversing the slide by flipping it through a central vertical axis, as represented in figure (1c). This case can be better understood by its similarity to the case of flipping 3D colour glasses in stereo photography. More experimental results and analytical calculations will be published elsewhere.

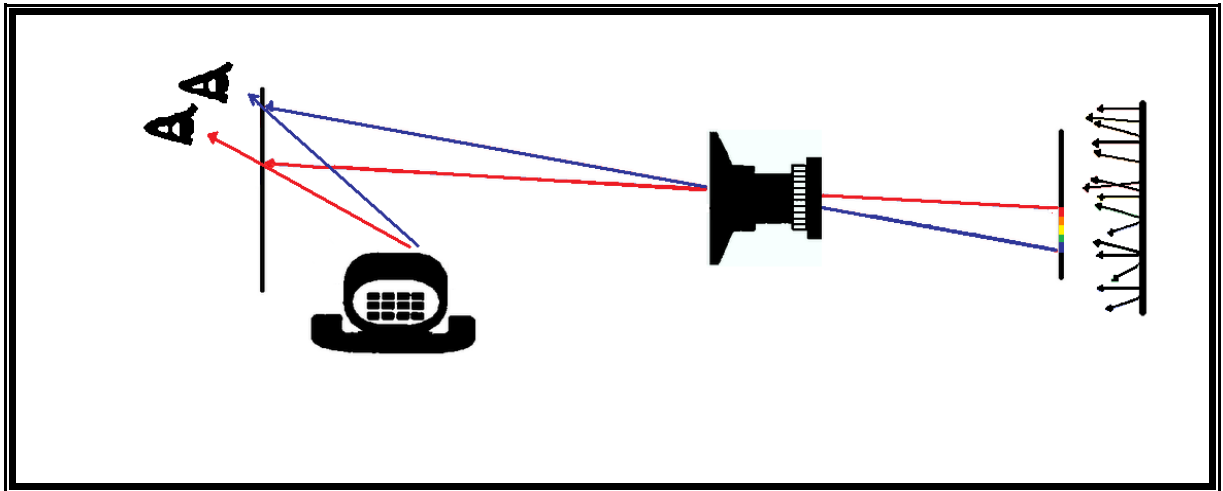
*Figure 1c : Inverse path reproduction process II*

**Obtaining an Enlarged Image**

An enlarged image may be obtained if a diffractive screen is used as in Lunazzi (3)(4) constituting an interesting case of enlarged holo–images (images that may keep the continuous horizontal parallax). This case can be easily interpreted by considering that the sequence of views focused at the diffractive screen allows for observing a single view from each single point of view, the one that corresponds to the wavelength that, being diffracted from the screen is precisely directed to the pupil of the observer. The focusing effect of the screen allows for this, while the wavelength diversity allows for the continuous horizontal parallax. We show in figure (2a) how two images points are reconstructed in front of the screen.

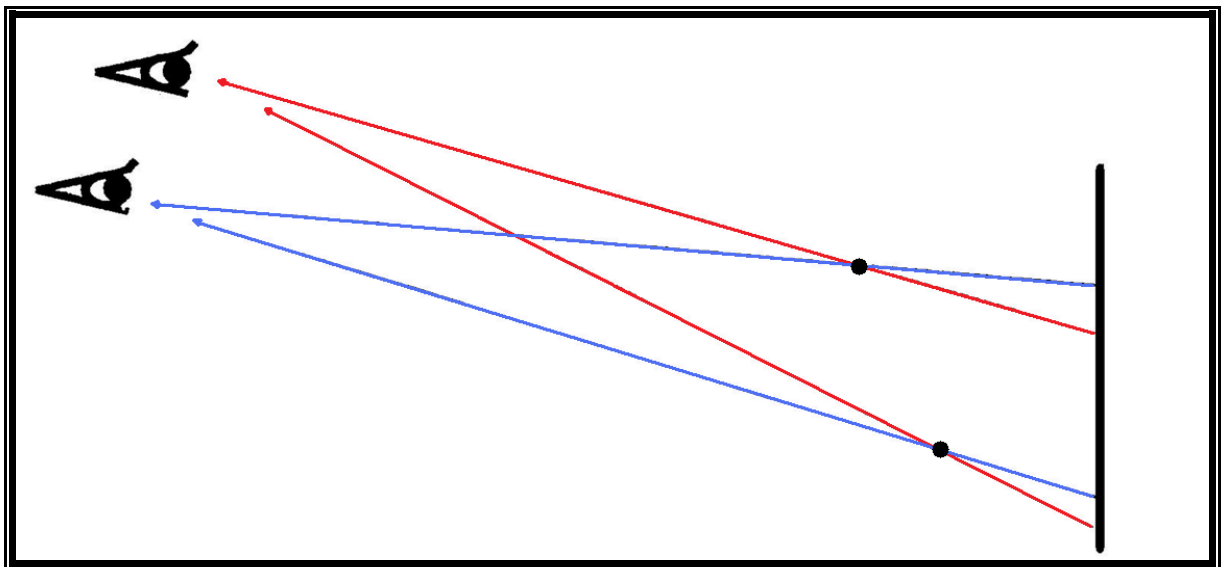
*Figure 2a : Representation of two 3D image points by means of a diffractive screen.*

The quality of this process depends mainly on the wavelength – reproduction capacity of the photographic process, as Lunazzi (5) demonstrated by substituting the registering process for a direct projection process, obtaining high quality and many centimetres deep images.

**The Wavelength – Reproduction Problem**

One limitation appears at the moment of photographing, since for deep scene large focal depth is needed, and a very reduced aperture must be used, so increasing exposure times. A more important limitation arises from the non–physical condition which is derived from conventional colour photography. The three–chromatic procedure is reasonably good for direct visual applications but does not allow to give to a traversing ray of white light the wavelength of the original incident ray. The transmitted ray receives a large wavelength band corresponding to the transmission property of one or up to three dye films on the photograph. Light that should be monochromatic exists the film with more than one third of the visible spectral range. As a first reasoning, we interpreted that the more reduced bandwidths correspond to those wavelength that more closely match the peak values of the sensibility of the dye. If we consider that the interval between the extreme wavelengths (red–blue) should be registered as a sequence of sharp points with a single wavelength value, the actual situation involves the joining of a many points whose bandwidth has common wavelength value. The result of this is that what can be seen from one point of view is an extended segment whose length is about one third of the length separating the extreme registered points (red–blue). We show in figure (2b) a ray tracing scheme for the problem. We considered that only an intermediate image point becomes diffracted (shadowed region on the figure) due to its extended bandwidth. The sharpness of the image and location of points in depth becomes clearly disturbed.

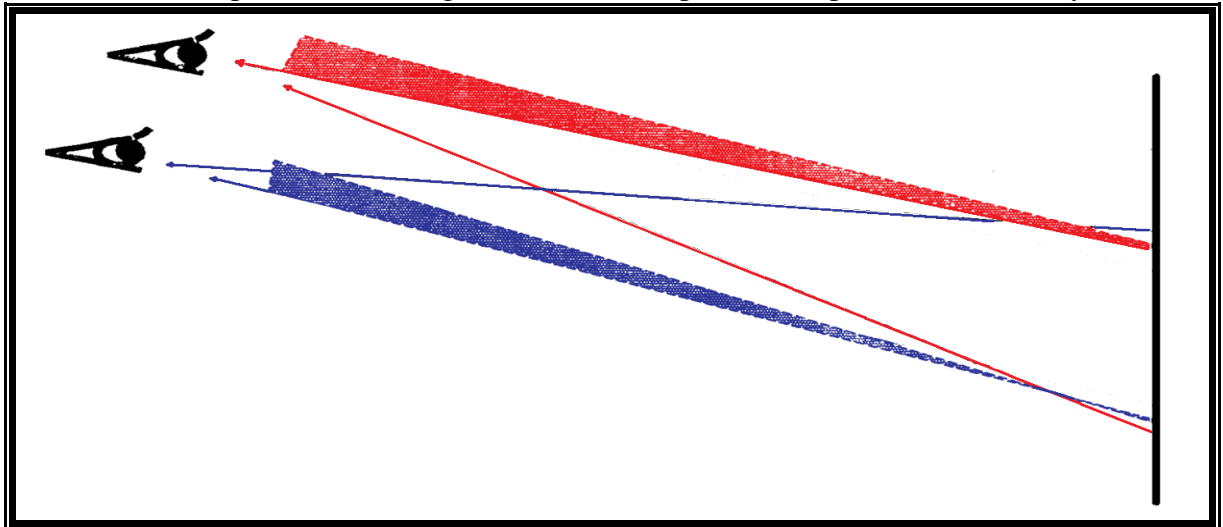

Figure 2b : Mixing of rays due to insufficient chromatic selectivity.

It is surprising to observe that, even so, the photographs may give a representation of depth, limited to the case of objects that are not very far from the plane of the diffraction grating. From one extreme to another of the observation region, scenes appear differentiated. While no discontinuous transition is noticeable when looking around. This technique, we named "holophotography" and, although the registration stage could be replaced by a non–diffractive process (i.e. photographing through an interference filter with continuous varying value of wavelength transmission) the combination between the two diffraction complementary processes gives an intrinsic quality to it. Since holography resulted from the combination of an interference and diffraction process, it is interesting to search for new combinations of these processes. There is an interference process that could be included in our double–diffraction process to solve the problem of wavelength selectivity, which is Lippmann photography (6). this technique allows, in theory, for the true reproduction of the wavelength value and was demonstrated by Fleisher et al (7) to allow for the superposition of ten photographed images being individually separated by the chromatic selection of interference filters. Our experimental results stimulate us to obtain nice sharp deep images through a process which would then combinate diffraction encoding, interferencial registering and diffractive reconstruction.

**Acknowledgements**

The author acknowledges financial help from the National Council for Research – CNPq and from the Found for Assistance to Research FAEP – UNICAMP.